# CEPC Input to the ESPP 2018 - Physics and Detector

CEPC Physics-Detector Study Group

## Abstract

The Higgs boson, discovered in 2012 by the ATLAS and CMS Collaborations at the Large Hadron Collider (LHC), plays a central role in the Standard Model. Measuring its properties precisely will advance our understandings of some of the most important questions in particle physics, such as the naturalness of the electroweak scale and the nature of the electroweak phase transition. The Higgs boson could also be a window for exploring new physics, such as dark matter and its associated dark sector, heavy sterile neutrino, et al. The Circular Electron Positron Collider (CEPC), proposed by the Chinese High Energy community in 2012, is designed to run at a center-of-mass energy of 240 GeV as a Higgs factory. With about one million Higgs bosons produced, many of the major Higgs boson couplings can be measured with precisions about one order of magnitude better than those achievable at the High Luminosity-LHC. The CEPC is also designed to run at the Z-pole and the W pair production threshold, creating close to one trillion Z bosons and 100 million W bosons. It is projected to improve the precisions of many of the electroweak observables by about one order of magnitude or more. These measurements are complementary to the Higgs boson coupling measurements. The CEPC also offers excellent opportunities for searching for rare decays of the Higgs, W, and Z bosons. The large quantities of bottom-quarks, charm-quarks, and tau leptons produced from the decays of the Z bosons are interesting for flavor physics. The clean collision environment also makes the CEPC an ideal facility to perform precision QCD measurements. Several detector concepts have been proposed for the CEPC. Dedicated simulation and R&D program confirm these concepts can fulfill the CEPC physics requirements.

In this document, we provide a brief summary of the physics potential and the detector design concepts, both of which are laid out in detail in the Conceptual Design Report (CDR) released in November 2018. We also outline future directions and challenges. In the Addendum, we briefly describe the planning and the international organization of the CEPC. The next step for the CEPC team is to perform detailed technical design studies. Effective international collaboration would be crucial at this stage. This submission for consideration by the ESPP is part of our dedicated effort in seeking international collaboration and support. Given the importance of the precision Higgs boson measurements, the ongoing CEPC activities do not diminish our interests in participating in the international collaborations of other future electron-positron collider based Higgs factories.

# Introduction

The historic discovery of the Higgs boson in 2012 by the ATLAS and CMS Collaborations at the Large Hadron Collider (LHC) at CERN completes the particle spectrum of the Standard Model (SM). Responsible for the masses of elementary particles, the Higgs boson is at the heart of the SM. It is connected to many important open questions in the particle physics, such as the large hierarchy between the weak and the gravitational interactions, the nature of the electroweak phase transition, and its role in the early universe. Moreover, the Higgs boson could also be a window to the new physics, like the dark matter and the associated dark sector.

For the foreseeable future, the LHC is the place-to-be to study the Higgs boson. The planned high luminosity upgrade of the LHC (HL-LHC) will produce over 100 million Higgs bosons. However, the precisions of the Higgs boson property measurements at the HL-LHC are limited by large theoretical and experimental uncertainties. The ultimate relative precisions achievable for the majority of the Higgs boson couplings are estimated to be around 5%-10%.

Compared with the LHC, electron-positron colliders have significant advantages for the Higgs boson property measurements. The Higgs boson signal is largely free of QCD backgrounds and the signal to background ratio is significantly higher. Moreover, Higgs boson candidates can be identified through the recoil mass method without tagging the Higgs boson decay products, allowing for the measurements of the Higgs boson width and couplings in a model-independent way. Multiple Higgs factories based on electron-positron colliders have been proposed, including the International Linear Collider (ILC), the Compact Linear Collider (CLIC), the Future Circular Collider (FCC-ee), and the Circular Electron Positron Collider (CEPC).

The CEPC was proposed by the Chinese high energy physics community soon after the Higgs boson discovery and is expected to be hosted in China. It has a main ring with a circumference of 100 km. The CEPC is designed to operate at several center-of-mass energies, including 91.2 GeV as a Z factory, 160 GeV at around the W pair production threshold, and 240 GeV as a Higgs factory. In its planned ten-year operation with two detectors, the CEPC will deliver combined total integrated luminosities of 16, 2.6, and 5.6 ab$^{-1}$ for the Z, the W, and the Higgs operation, respectively. It will produce close to one trillion Z bosons, 100 million W bosons, and over one million Higgs bosons. Billions of bottom quarks, charm quarks, and tau leptons will also be produced, mainly through Z boson decays, making the CEPC a de-facto B-factory and a tau-charm factory.

The CEPC is also a synchrotron light source that can serve wider scientific communities. After the completion of the CEPC physics program, a proton-proton collider with a center-of-mass energy of 100 TeV (the Super proton-proton collider, or SppC) could be installed in the same tunnel, allowing proton-proton and heavy ion collisions.

We submit this document on the CEPC physics potential and detector design for the consideration by the ESPP. More detailed information has been presented in the CEPC Conceptual Design Report [1]. We believe an electron-positron collider based Higgs factory is an indispensable next step for the exploration of the Higgs boson physics. The CEPC project has the potential to characterize the Higgs boson in the same way LEP did to the Z boson and to search for the possible deviations from the Standard Model. Aiming for precision measurements as a gateway to advance our understandings of some of the most important questions in particle physics, the project fits well in the strategy of the international particle physics community. We would like to seek the support from the ESPP for active research programs in Europe and elsewhere towards the realization of electron-positron collider based Higgs factories such as the CEPC.

## The Physics Case

Table 1 presents the nominal parameters of the CEPC [2] and the expected boson yields [2]. As these numbers suggest, the CEPC is not only a Higgs factory but also factories for W and Z bosons. Numerous b-quarks, c-quarks, and tau-leptons will also be produced from the decays of the Higgs, W, and Z bosons, offering excellent opportunities for studying flavor physics.

| Operation mode | Z factory | W threshold scan | Higgs factory |
|---|---|---|---|
| Ecm (GeV) | ~91.2 | 158 – 172 | 240 |
| L($10^{34}$cm$^{-2}$s$^{-1}$) | 17 – 32 | 10 | 3 |
| Running time (years) | 2 | 1 | 7 |
| Integrated luminosity (ab$^{-1}$) | 8 – 16 | 2.6 | 5.6 |
| Higgs yield | - | - | $10^6$ |
| W yield | - | $10^7$ | $10^8$ |
| Z yield | $10^{11\text{-}12}$ | $10^9$ | $10^9$ |

Table 1: The nominal parameters and boson yields of the CEPC [2].

To maximize the number of the Higgs bosons produced while keeping the synchrotron radiation at a manageable level, the CEPC will operate at a center-of-mass energy of 240 GeV for its Higgs operation. The typical event rate of the SM processes within the detector acceptance is 10 Hz with one Higgs boson produced every 2 minutes. The CEPC detectors can record almost all these events. The detectors are required to efficiently distinguish the Higgs boson signal from backgrounds, and identify different production/decay modes of the Higgs boson.

At the CEPC, the Higgs bosons are dominantly produced in association with the Z bosons (ZH events). Because the energy of electron-positron collisions is precisely known, the ZH events can be identified from the recoil mass information using the visible Z boson decays, see Figure 1 for example. This allows for the measurement of the ZH production cross section independent of the Higgs boson decay.

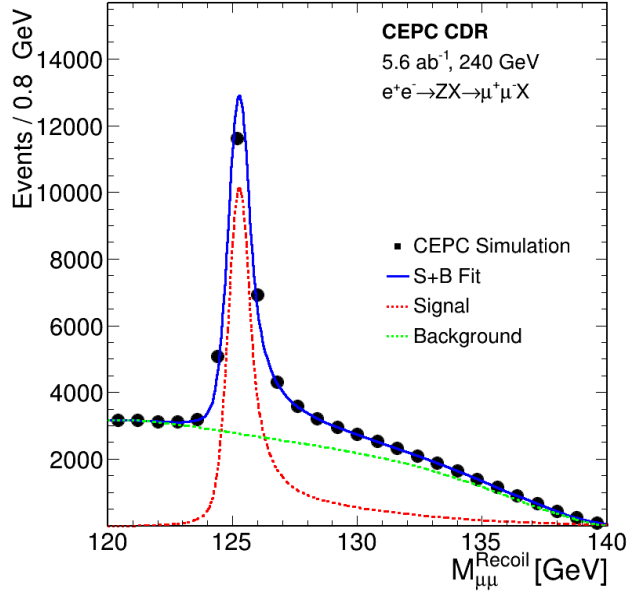

Figure 1: The invariant mass distribution of the system recoiling against a pair of muons from the CEPC Higgs operation. A resonance structure at the Higgs boson mass is expected from ZH events with the Z→μμ decay while a continuum distribution is predicted from background events [1].

| Property | Estimated Precision |
|---|---|
| $m_H$ | 5.9 MeV |
| $\Gamma_H$ | 3.1% |
| $\sigma(ZH)$ | 0.5% |
| $\sigma(\nu\nu H)$ | 3.2% |

| Decay mode | $\sigma(ZH)\times BR$ | BR |
|---|---|---|
| H→bb | 0.27% | 0.56% |
| H→cc | 3.3% | 3.3% |
| H→gg | 1.3% | 1.4% |
| H→WW | 1.0% | 1.1% |
| H→ZZ | 5.1% | 5.1% |
| H→γγ | 6.8% | 6.9% |
| H→Zγ | 15% | 15% |
| H→ττ | 0.8% | 1.0% |
| H→μμ | 17% | 17% |
| H→inv | – | <0.30% |

Table 2: Projected precisions of the CEPC measurements of the properties of the Higgs boson [1][3]. All precisions are relative except for $m_H$ and Br(H→inv) for which $\Delta m_H$ and the 95% CL upper limit on BSM decays are quoted respectively.

The Higgs boson rates (the production cross section times the decay branching ratio, or σ×BR) can be measured with relative precisions of percent or sub-percent for its main decay modes (bb, cc, gg, WW, ττ) and between 10%-20% for its rare decay modes (γγ, μμ, Zγ). Moreover, the Higgs boson total width can be measured, model-independently, to a relative precision of 3%. More detailed projections are summarized in Table 2. These measurements can be used to extract relevant Higgs boson couplings. Figure 2 compares the relative precisions of the Higgs boson coupling measurements at the CEPC and the HL-LHC. The measurements of the Higgs boson couplings can reach precisions of 0.1%–1%, approximately one order of magnitude better than the expected precisions of the HL-LHC measurements.

The CEPC also has an excellent capability for the precise measurements of the electroweak observables. Left panel of Figure 3 lists the expected precisions from the CEPC compared with those from the LEP experiments. The right panel of Figure 3 shows the CEPC constraints on the electroweak oblique parameters in comparison with the current precisions. The CEPC can improve the current precisions of the electroweak measurements by at least one order of magnitude.

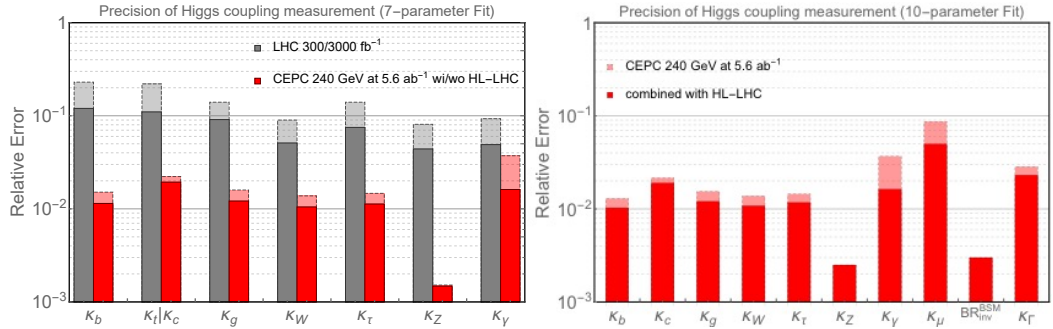

Figure 2: Expected relative precisions of the Higgs boson property measurements at the CEPC [1][3], with an integrated luminosity of 5.6 ab$^{-1}$ at $E_{CM}$ = 240 GeV, and the comparison with the LHC/HL-LHC. In the left panel, a constrained 7 parameter fit (chosen to facilitate a comparison with the LHC) in the κ-framework is presented. In the right panel, a model independent 10 parameter fit is presented.

The precision measurements at the CEPC allow us to make significant progress in addressing several of the most important open questions in particle physics, in particular those associated with the weak interaction.

Measurements of the Higgs boson properties can also help to reveal the dynamics that controls the nature of the electroweak phase transition. If the transition is of the first order, it opens up a great possibility of explaining the matter antimatter asymmetry in the Universe. The CEPC cannot directly probe the triple Higgs coupling. However, it is very sensitive to any new physics which significantly alters the dynamics of the electroweak phase transition since such new physics can modify other Higgs boson couplings. This is demonstrated in the singlet augmented model shown in Figure 5.

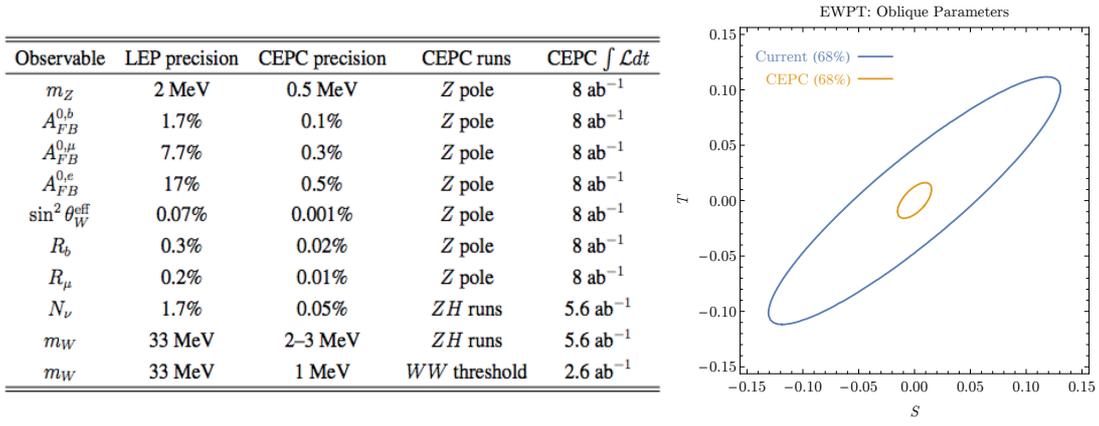

Figure 3: Left: The expected precisions in a selected set of EW precision measurements at the CEPC and the comparison of precisions with the LEP experiments [1]. Right: The constraint on the oblique parameters [1].

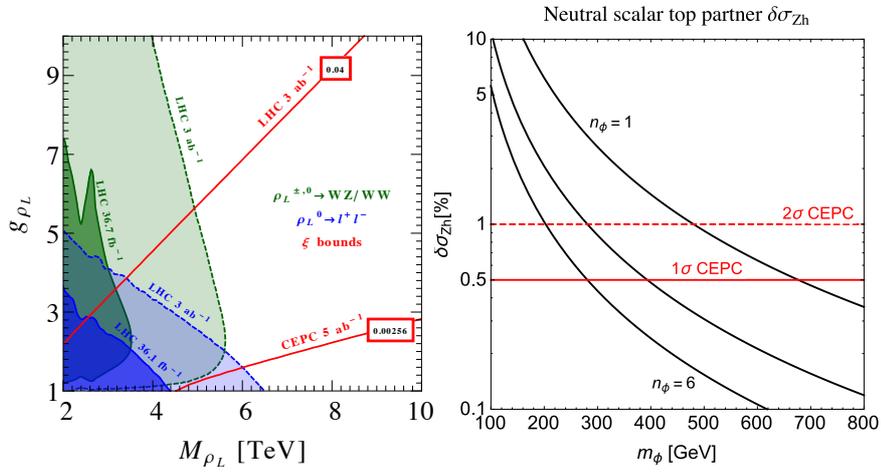

Figure 4: Testing naturalness with the Higgs coupling measurements in composite Higgs models (left) and models with neutral top partners (right) [1].

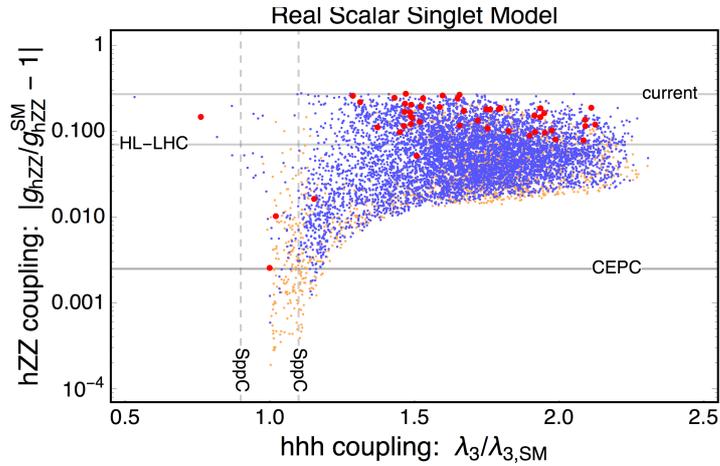

Figure 5: CEPC sensitivity, through the measurement of the Higgs-Z coupling, to models which have a first order electroweak phase transition [1].

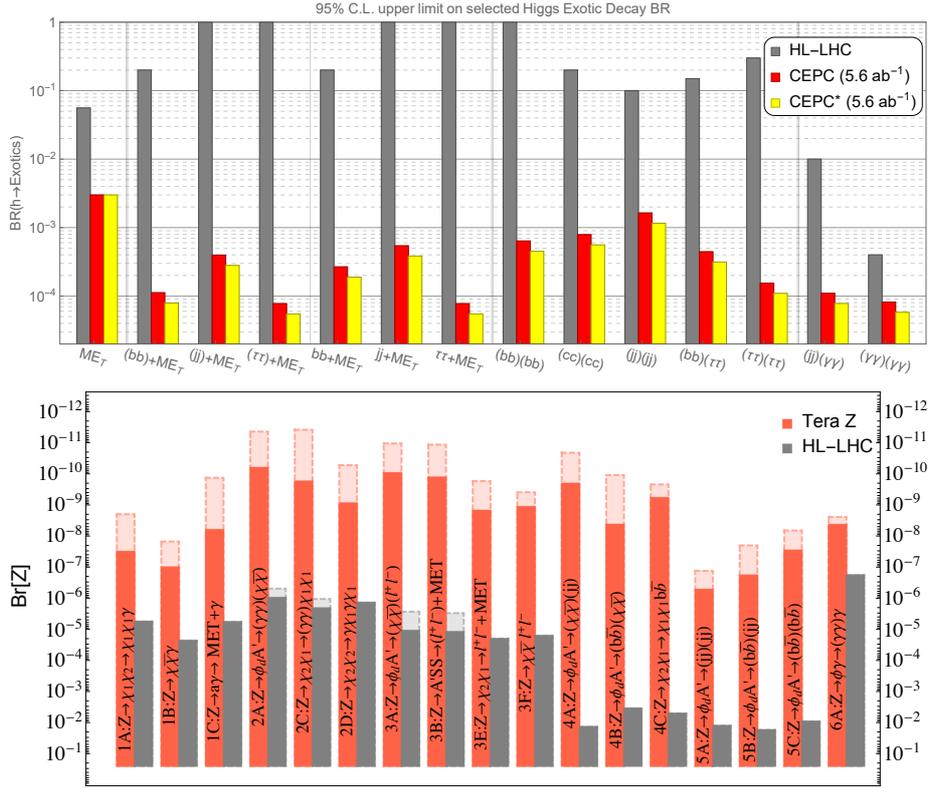

Figure 6: Reach for exotic Higgs boson decays at the CEPC (top) and reach for exotic Z decays at the CEPC (bottom) [1].

The CEPC is also an excellent place to search for new particles. The exotic decays of both the Higgs and the Z bosons provide very sensitive probes, as shown in Figure 6. The identification of the Higgs bosons via the recoil mass method together with its high reconstruction efficiency makes it extremely sensitive to probe the exotic decays of the Higgs bosons at the CEPC. Beyond SM (BSM) decays can be probed for branching ratios down to $10^{-3}$ to $10^{-5}$. Moreover, the Higgs boson width provides an inclusive limit on all BSM decays. This is particularly interesting given the unique capability of the Higgs boson coupling to dark matter and other dark sector states. With the huge statistics of Z bosons, the sensitivity of many rare Z decay channels can be significantly enhanced in comparison with the HL-LHC. Rare Z boson decays also offer an excellent opportunity to search for heavy sterile neutrinos.

Last but not least, the CEPC offers many opportunities for flavor physics and tau physics, as large quantities of b-quarks, c-quarks, and tau-leptons will be produced from the decays of the Higgs, W, and Z bosons. In particular, the decays of nearly $10^{12}$ Z bosons will result in approximately $10^{11}$ B hadrons, comparable to the number expected from the super-B factory. A similar number of Charm hadrons will be produced as well. The CEPC can discover and study heavy B hadrons which are inaccessible at B factories or are too complex to be identified at the LHCb. Thus the CEPC offers complementary flavor physics program. The CEPC is also a great facility for precision QCD measurements. It will offer a new opportunity for determining the QCD coupling strength through event shape studies. In addition, the clean environment reduces both experimental and theoretical uncertainties. It is an ideal place to study a number of subtle QCD effects.

Looking forward, several challenges remain to be addressed with additional physics studies. In particular, the understanding of the CEPC's potential for flavor physics and QCD measurements are still preliminary. A more comprehensive survey, supported by simulations, is needed to obtain more concrete estimates. In addition, the expected small statistical uncertainty at the CEPC makes stringent requirements on reducing the theoretical uncertainty and calls for the implementation of state-of-the-art theoretical calculations in Monte Carlo simulation tools.

# The Detector Requirements and Concepts

The CEPC detector performance requirements are driven by the physics cases described above. The CEPC detectors must be able to identify and measure all key physics objects with high efficiency, purity, and precision. Charged particles and unconverted photons with energy above 1 GeV within the detector fiducial region are required to be reconstructed with an efficiency higher than 99%. Isolated leptons with a momentum larger than 5 GeV must be identified with an efficiency greater than 99% and a mis-identification rate from hadrons less than 1%. A di-jet mass resolution of better than 4% is required to separate the hadronic decays of the Higgs, W, and Z bosons. Flavor tagging of jets is essential for disentangling different Higgs boson decay modes and for EW measurements. Benchmarked using a sample of inclusive Z→qq events at 91.2 GeV, efficiency/purity better than 80%/80% for the b-jets and better than 60%/60% for the c-jets are required.

The CEPC physics program also requires a precise determination of the luminosity. For the Higgs factory (Z factory) operation, the luminosity must be determined with a relative accuracy of $10^{-3}$ ($10^{-4}$). The precise knowledge of the beam energy is critical for the precise Higgs boson mass and EW measurements. The beam energy scale needs to be measured with an accuracy of 1 MeV for the Higgs factory operation, and 100 keV for the Z factory and W threshold scan operations.

Two interaction points are envisioned for the CEPC. In the recently completed CDR, two independent detector concepts have been proposed. The baseline concept is a particle-flow oriented detector design that uses ultra-high granularity calorimeter and 3-Tesla solenoid. An alternative detector concept is based on the dual-readout calorimeter technology, large radius drift chamber, and 2-Tesla solenoid.

The baseline detector concept is developed from the International Large detector (ILD) through a series of optimization for the CEPC. To accommodate the CEPC final focusing system, the machine detector interface and the forward region have been redesigned. The detector, as illustrated in Figure 7, is composed of a high precision silicon based vertex and tracking system, a Time Projection Chamber (TPC), a silicon-tungsten sampling electromagnetic calorimeter (ECAL), a resistive plate chamber (RPC)-steel sampling hadron calorimeter (HCAL), a 3-Tesla solenoid, and a muon/yoke system. A baseline reconstruction software toolkit has been developed. It has been demonstrated using the full simulation that the baseline detector concept can meet the requirements set forth above. In the barrel region, the track momenta can be measured with a relative accuracy of O(0.1%). For isolated leptons with energy larger than 2 GeV, the identification efficiency is better than 99.5% with a mis-identification rate of <1%. The rate of jets misidentified as photons is found to be negligible. For hadronic decaying tau leptons, the identification efficiency approaches 90% with a percent-level mis-identification rate. By employing the particle flow algorithm, a jet energy resolution of 3.5%-5.5% is achieved for jets with energy of 20 GeV to 100 GeV. The mass resolution reaches 3.8% for the hadronic decays of the Higgs, W, and Z bosons, see Figure 8. The jet flavor tagging has also been shown to reach the required performance [1, 4].

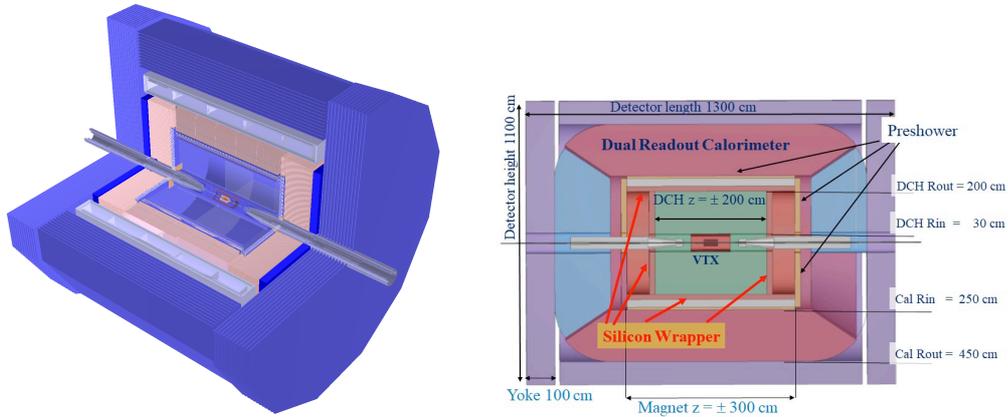

Figure 7: The baseline (left) and alternative (right) detector concepts.

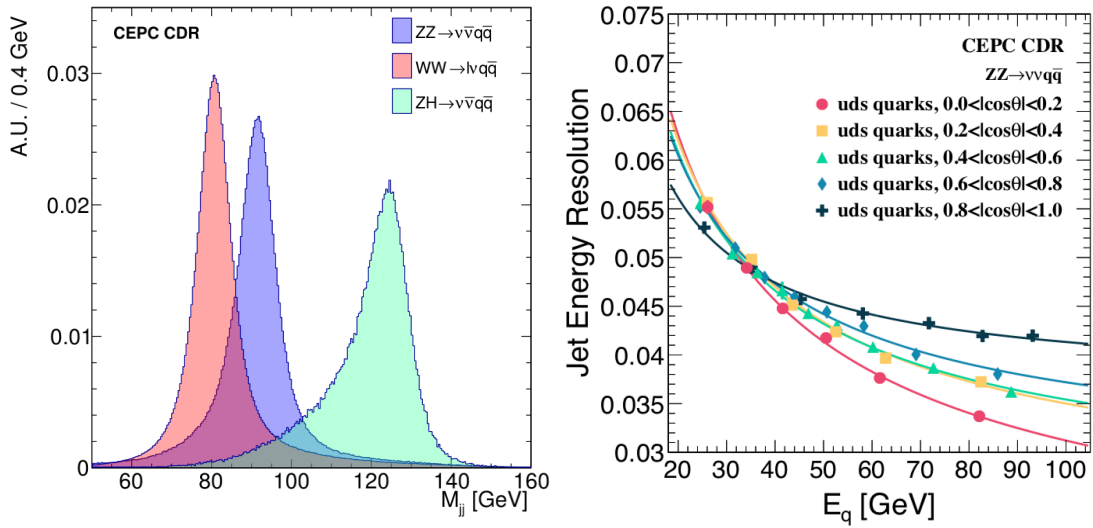

Figure 8: The W, Z, H boson mass reconstructed with corresponding 2-jet events (left) and jet energy resolutions as functions of jet energy (right) [1].

The CEPC sensitivity to some selected physics processes has been demonstrated with the full detector simulation. For the alternative detector concept, simulation and reconstruction tests have only been performed at sub-detector level, but its overall performance is expected to be similar to that of the baseline concept.

For both the baseline and the alternative detector concepts, intensive R&D programs have been launched. International partnerships and collaborations are established between the CEPC study group and multiple international institutes. Members of the CEPC study group are active in many detector R&D programs, including LC-TPC and CALICE. Multiple international institutes have a signed the Memorandum of Understanding on the CEPC collaboration. Approximately 280 foreign researchers from 140 different international institutes and universities participated in the studies for the CEPC CDR. Critical challenges have been identified and will be addressed with dedicated R&D and simulation studies. The list includes:

- Continued optimization of the detector design: balancing the performance and cost, taking advantage of the latest development in technology.
- The global integration of the detector including mechanical, thermal, and electrical.
- The systematic control and in-situ monitoring. The stability of the CEPC detector system is critical for precision measurements, particularly at the Z pole operation.
- Advanced reconstruction algorithm development, computing and data storages.

## Messages for the Strategy Process

In 2018, the CEPC studies have accomplished the critical milestone of the Conceptual Design Report [1, 2]. These collective studies have converged in an accelerator design with luminosity goals for different CEPC operation modes, and several conceptual detector designs that should meet the CEPC physics requirements. The joint theory, detector design, and simulation studies have given a clear demonstration of the CEPC physics capabilities: the CEPC has an immense potential towards the precision Higgs and EW measurements, and can provide complementary information in flavor physics.

With relevance to the European Strategy process, two clear messages stand out for the next steps of the CEPC project:

A comprehensive detector design optimization is needed to maximize the physics potential and reduce cost. The ongoing and planned detector R&D will address the main technical challenges. The ultimate CEPC physics reach, especially for the precision EW measurements, is largely dominated by the experimental systematic and theoretical uncertainties. The CEPC potential for flavor and QCD physics needs to be better quantified. The technical design of the CEPC detector and the theory-phenomenology studies should address these issues.

Although the CEPC already includes active international participation, a stronger international collaboration is crucial for the technical design studies and the eventually realization of the CEPC project. The CEPC study group is seeking to form an effective international organization for the CEPC project. While proposing the CEPC, the CEPC study group will strongly support and actively participate in any future electron-positron Higgs factories.

# Addendum 1: The Planning of CEPC

The Chinese government has established a program to host China initiated international large science projects. A call for proposals is expected in early 2019. After a selection process, the Ministry of Science and Technology will cultivate 3-5 seed projects by 2020, from which 1-2 projects will be approved for construction. China will continue to identify and support further large science projects in the future. This program provides a natural path for the realization of the CEPC.

The current priority of the CEPC project is to secure its position as one of the seed projects of this program. The official approval of the CEPC project, at the earliest, could happen in 2022. The accelerator construction could start soon after the approval. At the same time, a call for detector Letters of Intent is planned. Two detectors will be selected and the International collaborations will be formed accordingly. The collaborations should deliver their Detector Technical Design Reports within two years after the starting of the construction of the accelerator.

In November 2018, the CEPC Conceptual Design Report was released and the CEPC International Advisory Committee provided its recommendations. According to which, the accelerator study group should complete its Technical Design Report by 2022. Meanwhile, collaborative R&D work on each of sub-detector systems will be the focus. To achieve this goal, two International Committees will be established. The first is an Accelerator Review Committee that advises on all matters related to the accelerator design and R&D including the Machine-Detector Interface and upgrade capabilities. The second is a Detector R&D committee that reviews and endorses the Detector R&D proposals from the international community, such that the international participants could apply for funds from their funding agencies and make effective and sustained contributions.

International collaboration is vital for the CEPC project. Active collaborations have been established between the domestic CEPC study groups and multiple international research institutions. Through these collaborations, many key challenges of the CEPC detector design and physics studies have been identified and being addressed through dedicated R&D programs.

A new organization structure is proposed to promote and accommodate the future international participation, see Figure 9. This structure reflects the discussion at the 2018 CEPC workshop and takes into account the recommendations of the International Advisory Committee. It is intended for the period from 2019 till the construction.

In this structure, all the building blocks will integrate the international participation. The Institution Committee writes the bylaws and makes major decisions on organizational issues. The national representatives interface with the National Funding Agencies and the corresponding institutions are represented in the Institution Committee. Supported by the Accelerator Review Committee and the Detector R&D Committee, the Project director is responsible for the coordination of studies at each group.

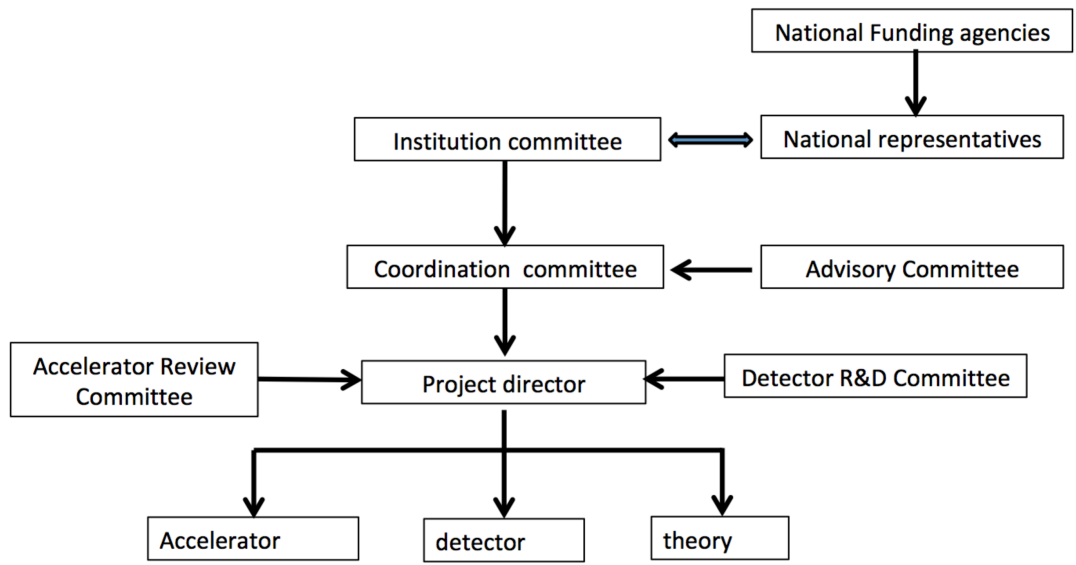

Figure 9, The planned international organization from 2019 till the construction

The organization will evolve with time. In the construction and operational phase, the organization will naturally evolve to include the council representing the participating countries and the host lab management who provide supervision to this project.

Though the CEPC project is initiated by and to be hosted in China, it is envisioned to be an international project. The organization and the management of the project will reflect the international participation of its stakeholders. The successful international participation played a critical role in the delivery of the CEPC conceptual design, and it will certainly become more important in the future.